# Combined coherent x-ray micro-diffraction and local mechanical loading on copper nanocrystals.

G Beutier[1], M Verdier[1], M de Boissieu[1], B Gilles[1] and F Livet[1]
[1] SIMaP, CNRS – Grenoble-INP – UJF, Saint-Martin d'Hères, France

M-I Richard[2,3], T W Cornelius[2,3], S Labat[2] and O Thomas[2]
[2] IM2NP, CNRS – Aix-Marseille University, Marseille, France
[3] European Synchrotron Radiation Facility, Grenoble, France

E-mail: guillaume.beutier@simap.grenoble-inp.fr

**Abstract.** Coherent x-ray micro-diffraction and local mechanical loading can be combined to investigate the mechanical deformation in crystalline nanostructures. Here we present measurements of plastic deformation in a copper crystal of sub-micron size obtained by loading the sample with an Atomic Force Microscopy tip. The appearance of sharp features in the diffraction pattern, while conserving its global shape, is attributed to crystal defects induced by the tip.

## 1. Introduction
The recent excitement about size effects on mechanical properties of crystals [1] still lacks a good understanding. It is now clear that the crystal microstructure (strain and defects) plays an even greater role at smaller scale than at larger scale [2], but its detailed influence remains unexplained. There is thus a need to develop a non invasive probe that could measure the microstructure of a crystal and its evolution under mechanical loading. Coherent X-ray Diffraction (CXD) could be such a probe: its capability to image the crystal strain field in three dimensions has been demonstrated [3], and it also has a great sensitivity to crystal defects [4,5], although imaging in the presence of defects remains an unsolved problem. On one hand, when a crystal is defect free, its diffraction pattern around a Bragg reflection is smooth, except for possible fringes due to its shape. The global shape of the diffraction pattern is given by the strain field inhomogeneity. On the other hand, crystal defects such as dislocations create sudden phase jumps in the scattering amplitude between both sides of the defect, inducing sharp features in the diffraction pattern [4,5]. In weakly strained crystals, they can sometimes be identified by the splitting of a chosen Bragg peak [5]. The case of highly strained crystals, such as the nanocrystals studied in the present paper, is more favourable to the observation of crystal defects thanks to the larger portion of the reciprocal space covered by a significant scattered amplitude: phase jumps may be observed away from the centre of the Bragg peak. Recent experimental developments allow the combined use of microfocused synchrotron radiation with Atomic Force Microscopy (AFM) [6]. Here we use this experimental set-up to measure CXD while *in situ* loading a model system for the study of plasticity: defect-free copper nanocrystals.

## 2. Experimental set-up

Copper islands were prepared by solid state dewetting of a 5 nm thin copper film on an atomically flat tantalum (001) substrate, as described in [7]. Most of them reach close-to-the-equilibrium shapes during the growth: truncated pyramids with a (001) top facet and (111) side facets. The square basis is around 800 nm wide and the height around 300 nm, with a small dispersion in size. They are separated from each other by a few microns. They grow in orientation relationship (001)[110]//(001)[100] with the tantalum (001) substrate. The epitaxial relationship results in an inhomogeneous strain distribution with an average compression of 0.65% along the growth axis. Due to the growth process, they naturally form with very few crystal defects. Transmission Electron Microscopy confirms the perfect crystalline quality of the islands. They are thus model systems to study the first stages of plasticity.

Coherent x-ray diffraction was performed at beamline ID01 of the European Synchrotron Radiation Facility. A coherent portion of the monochromatic (8 keV) beam was selected with high precision slits by matching their horizontal and vertical gaps with the transverse coherence lengths of the beamline (20 μm and 60 μm respectively close to the sample position). The coherent beam was then focused to ~0.5 μm x ~0.3 μm using Fresnel Zone Plate, in order to illuminate a single copper island. The 002 Bragg reflection was measured with a Maxipix area detector (256x256 pixels of 55 μm) placed at 0.88 m from the sample, such that the finest diffraction features were largely oversampled. More details about the set-up are given in [7] and a discussion on the wavefront can be found in [8]. A blunt tungsten AFM tip with a radius of curvature of ~2 μm was used to apply stress *in situ* on a selected island. The tip was mounted on a quartz tuning fork, whose resonance frequency (31.9442 kHz) was monitored to approach the sample surface or stay at constant height over the sample. In the experiment reported here, the height of the tip was driven by a motor with 100 nm step size, thus much too large to finely control the deformation of the sample. The mechanical compliance of the whole system supporting the tip was also not precisely known (between 1-10 kN/m), and it was therefore impossible to determine the precise force and displacement of the tip into the sample. A new set-up is currently being developed in order to address these issues. However, the configuration of the sample with many similar islands all well orientated allowed to test many of them. Here we present the case of one island for which we believe we witnessed the first stages of plasticity.

One major difficulty of this type of experiments is to align together the x-ray beam (~0.5 μm x ~0.3 μm), the sample of interest (~0.8 μm x ~0.8 μm x ~0.3 μm) and the tip (~2 μm x 2 μm). To achieve this alignment, the sample and the tip were mounted on independent stacks of piezo-motors, both mounted on the regular sample stage of the diffractometer. Both stacks were scanned independently, in Bragg geometry at the copper 002 reflection: first the sample stack to obtain a map of the copper islands and precisely position one of them in the beam, then the tip to obtain a map of its "shadow" as the tip blocked either the incident or the reflected beam from the chosen copper island [9]. This method (aligning both the sample and the tip with respect to the beam) was found more reliable than aligning the tip with respect to the sample, since the x-ray beam is the smaller of the three elements. In particular, the shape of the tip prevented from imaging the sample surface with a good resolution, as it was just enough to locate and resolve individual copper islands.

While the 002 Bragg reflection of the selected island could be mapped in the three-dimensional reciprocal space by either rocking the sample or scanning the x-ray energy [10], a single slice of the reciprocal space obtained on the area detector without moving any motor is enough to observe the changes of the diffraction pattern as the sample is mechanically loaded with the tip.

## 3. Results

We present here results observed during a mechanical load on an island. A similar behaviour to that described below was observed on other islands.

Snapshots of the 002 Bragg reflection of the chosen copper island are taken while the tip is lowered into contact with the surface and beyond (figure 1). The initial diffraction pattern (figure 1a), before starting the compression, is typical of the 002 Bragg reflection for these copper islands [7]: the large inhomogeneous strain field induced by the epitaxial interface with the substrate dominates the diffraction pattern; it broadens the Bragg spot into an irregular shape; fringes bent by the strain field inhomogeneity and thick shoulders related to the {111} facets can be seen. This shape can be modelled with the Finite Element Method (assuming a pure thermoelastic behaviour) combined with a kinematic diffraction model based on the Fast Fourier Transform [7]: for the 002 Bragg reflection it is directly related with the displacement field $u_z$ along the growth axis. In the case presented here, the azimuth of the crystallite around the growth axis is $\psi=15°$, with the reference value 0° when a [111] direction of the island is in the scattering plane. A few features close to the maximum of intensity suggest that a couple of crystal defects are already present in the sample. The tip is then engaged into contact using the close-loop control based on the tuning fork frequency. The diffraction pattern measured with the tip engaged do not show any change. The z-motor of the tip is driven 200 nm further still without any visible change on the diffraction pattern (figure 1b). The intensity is notably conserved, showing that the tip does not shadow the incoming beam or the diffracted beam. The absence of change during these first 200 nm can be explained by the mechanical compliance of the AFM (between 1-10 kN/m). The next 100 nm step produces a sudden change in the diffraction pattern, with a loss of about half of the intensity (figure 1c): while the general shape of the pattern is roughly conserved, many speckles appear. They are the signature of crystal defects. The tip is then retracted to its initial position but the diffraction pattern (figure 1d) remains exactly as it was after indentation, and the intensity is not recovered: it confirms that the observed change is due to plastic deformation only. With the large sphere-like end of the AFM tip we use (~2μm radius), the initial loading is mainly elastic (Hertzian contact) until a burst of dislocations is nucleated [11]. In fact, no elastic deformation could be evidenced from the average position of the pattern on the detector during the compression and release. The displacement of intensity after loading toward the bottom of the detector (i.e. lower scattering angle) is consistent with the expected strain relaxation associated with the introduction of defects: the as-grown islands are compressed in the out-of-plane direction [7], such that strain relaxation increases the mean out-of-plane parameter and thus decreases the 002 Bragg angle. The comparison of the rocking curves of the reflection before and after loading (figure 2) confirms the reduction of the peak intensity and shows a broadening due to the apparition of crystal defects, such that the total intensity is almost conserved. The position of the peak is also slightly shifted, suggesting a rotation of the copper island, but numerical simulations show that the introduction of crystal defects can create this shift without any rotation of the sample [12]. The fact that a significant amount of intensity remains close to the copper 002 Bragg reflection, with a general shape similar to the initial shape, shows that the island after loading is still a well defined crystal with a similar shape and strain field. This statement is further supported by later investigations with Scanning Electron Microscopy, which did not reveal any severely damaged island, although we could not identify with certainty the loaded island.

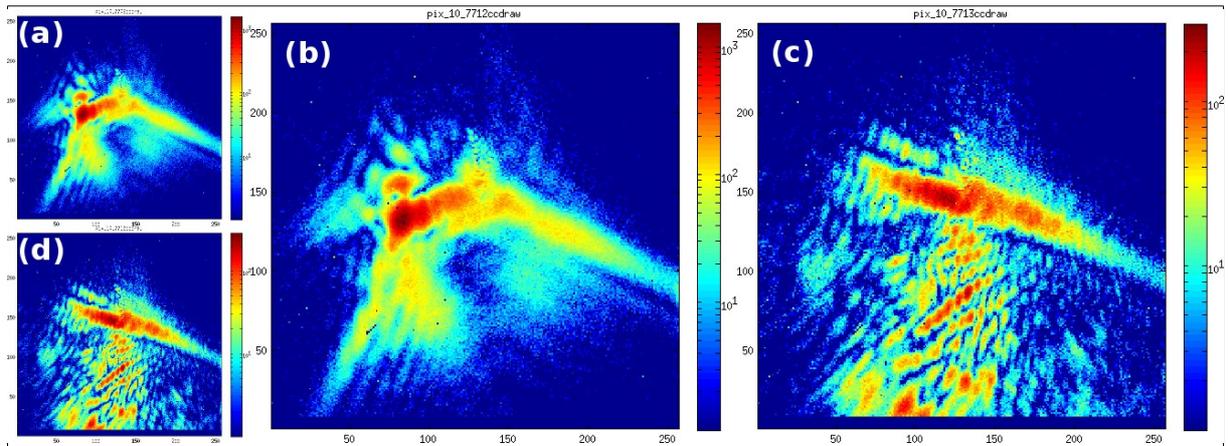

**Figure 1.** Full frame detector views of the 002 Bragg reflection at the top of the rocking curve during successive stages of loading; (a) tip retracted; (b) tip in contact; (c) tip displaced 100 nm beyond contact; (d) tip retracted. Acquisition time: 100 seconds. Logarithmic colour scale. The horizontal and vertical axes are respectively perpendicular and parallel to the scattering plane.

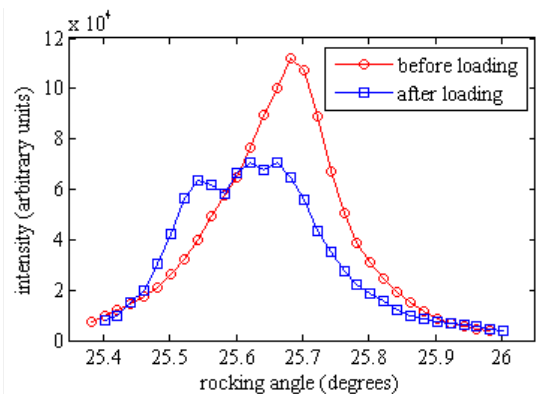

**Figure 2.** Rocking curve of the 002 Bragg reflection of the selected island before and after loading.

The nature of the crystal defects observed during this experiment remains unclear. Dislocations in a face-centred cubic crystal of lattice parameter $a$ have usually a Burgers vector of the type $\boldsymbol{b}=a<½,½,0>$ [13]. In a simple model, a dislocation introduces a sudden phase shift of $2\pi\boldsymbol{b}.\boldsymbol{q}$, with $\boldsymbol{q}$ the scattering vector, in the scattering amplitude. In the case of the 002 Bragg reflection, $\boldsymbol{b}.\boldsymbol{q}$=-1,0,1, depending of the particular Burgers vector. In any case it does not create a measurable phase shift. However, dislocations in copper are dissociated in two partial dislocations with Burgers vectors of type $a/6<1,1,2>$. The stacking fault in-between the two partials is thus phase-shifted by $2\pi/3$ with respect to the surrounding material at the 002 Bragg reflection. Nevertheless the dissociation extends only over a few nanometers (typically 2-5 nm) [13] and the proportion of phase-shifted volume in the copper island remains too small to create strong interference patterns, unless the density of dislocations becomes extremely high (~$10^{15}$ m$^{-2}$).

## 4. Concluding remarks

The combination of coherent x-ray micro-diffraction and *in situ* loading with an AFM tip allowed us to record the diffraction pattern of copper islands close to the 002 Bragg reflection during plastic deformation. High contrast speckles appear in the initially smooth diffraction pattern due to abrupt phase shifts in the scattering amplitude. These phase shifts are attributed to crystal defects whose nature remains unclear. In this experiment, the motors controlling the heights of the sample and of the tip were not sufficiently precise (10 nm for the sample height and 100 nm for the tip height) to gently load the crystals and introduce only a few dislocations. The applied force was not monitored, since the calibration is cumbersome with a tuning fork. A new set-up is being developed to improve the control of the load, in order to be able to observe the onset of plasticity: indeed, coherent x-ray micro-diffraction can evidence a single dislocation.

The authors acknowledge the European Synchrotron Radiation Facility for providing beamtime, and the staff of beamline ID01 for the support.